# $Ca_2Ru_{1-x}Cr_xO_4$ (0 < x < 0.13): Negative volume thermal expansion via orbital and magnetic orders

T.F. Qi[1,2], O.B. Korneta[1,2], S. Parkin[1,3], L. E. De Long[1,2], P. Schlottmann[4] and G. Cao[1,2*]

[1] Center for Advanced Materials, University of Kentucky

[2]Department of Physics and Astronomy, University of Kentucky

[3]Department of Chemistry, University of Kentucky

[4]Department of Physics, Florida State University

$Ca_2RuO_4$ undergoes a metal-insulator transition at $T_{MI}$ = 357 K, followed by a well-separated transition to antiferromagnetic order at $T_N$ = 110 K. Dilute Cr doping for Ru reduces the temperature of the orthorhombic distortion at $T_{MI}$ and induces ferromagnetic behavior at $T_C$. The lattice volume V of $Ca_2Ru_{1-x}Cr_xO_4$ (0 < x < 0.13) abruptly expands with cooling at both $T_{MI}$ and $T_C$, giving rise to a total volume expansion $\Delta V/V \approx 1$ %, which sharply contrasts the smooth temperature dependence of the few known examples of negative volume thermal expansion driven by anharmonic phonon modes. In addition, the near absence of volume thermal expansion between $T_C$ and $T_{MI}$ represents an Invar effect. The two phase transitions suggest an exotic ground state driven by an extraordinary coupling between spin, orbit and lattice degrees of freedom.

**PACs**: 65.40.De; 71.30.+h; 75.47.Lx

The layered 4d-transition metal oxides have attracted increasing interest since the comparable magnitudes of their intra-atomic Coulomb interaction U and 4d-bandwidth W can leave them precariously balanced on the border between metallic and insulating behavior, and/or on the verge of long-range magnetic order. Therefore, moderately strong spin-orbit interactions, as well as weaker perturbations such as slight changes in lattice parameters, can induce drastic changes in the character of their electronic ground states. This is dramatically illustrated by $Sr_2RuO_4$ and $Ca_2RuO_4$, where the prototypical p-wave superconducting state **[1]** of the former compound strongly contrasts with the more distorted structure (due to the smaller ionic radius $r_{Ca} < r_{Sr}$) and first-order metal-insulator (MI) transition observed for the latter **[2-3]**.

Extensive investigations of $Ca_2RuO_4$ **[4-5]** have established that a strong cooperative Jahn-Teller distortion removes the degeneracy of the three Ru $t_{2g}$ orbitals ($d_{xy}$, $d_{yz}$, $d_{zx}$) via a transition to ***orbital order*** that, in turn, drives the MI transition at $T_{MI}$ = 357 K **[6-12]**. Classic Mott insulators undergo ***simultaneous*** transitions to antiferromagnetic (AFM) order and an insulating state at $T_{MI}$. However, $Ca_2RuO_4$ undergoes AFM order at $T_N$ = 110 K << $T_{MI}$ **[2]**, and is therefore a highly interesting and unique archetype of a ***MI transition that is strongly coupled to a structural transition that is not driven by AFM exchange interactions***.

The MI transition of $Ca_2RuO_4$ is signaled by an abrupt jump in electrical resistivity ρ(T) accompanying a structural transition from a high-T tetragonal to low-T orthorhombic, which culminates in a striking increase of ρ(T) by over nine orders of magnitude between $T_{MI}$ and 70 K **[2-3]**, as shown in **Fig. 1a**. Although the **a**-axis of $Ca_2RuO_4$ contracts by 1.5% below $T_{MI}$, the **b**-axis expands by 3%; the combined effect of these conflicting uniaxial thermal expansions drives a particularly severe orthorhombic distortion of the tetragonal basal plane which shatters single-crystal samples and strongly contracts the lattice volume by 1.3% as T is lowered from 400 K to 70 K **[3]**.

Although controversy over the exact nature of the orbital ground state remains **[8-12]**, the extraordinary sensitivity of $T_{MI}$ and electrical resistivity to the application of modest pressure **[8]** or very slight changes in lattice parameters **[6-7]** clearly indicate that the lattice drives the highly unusual behavior of $Ca_2RuO_4$. On the other hand, Cr substitution for Ru in layered ruthenates generally results in no structural change, but does induce interesting magnetic behavior (as demonstrated by $SrRu_{1-x}Cr_xO_3$ **[13-15]**, $CaRu_{1-x}Cr_xO_3$ **[16]** and $Ca_3(Ru_{1-x}Cr_x)_2O_7$ **[17]**), which led us to pursue Cr substitution as an effective tool to investigate the curious dominance of lattice over magnetic degrees of freedom in $Ca_2RuO_4$.

Herein, we report the surprising finding that very dilute Cr doping for Ru in $Ca_2RuO_4$ not only induces spin canting and ferromagnetic (FM) behavior at a lower transition $T_C$ (which extrapolates to $T_N$ for x = 0), it strongly reduces the structural distortion and depresses $T_{MI}$. More important, ***the lattice volume V of $Ca_2Ru_{1-x}Cr_xO_4$ (0 < x < 0.12) abruptly expands with cooling just below $T_{MI}$ and, subsequently, just below $T_C$***, giving rise to a large net volume expansion $\Delta V/V \approx 1$ % as temperature decreases over the range 90 < T ≤ 295 K. The negative-volume-thermal-expansion (NVTE) we observe in $Ca_2Ru_{1-x}Cr_xO_4$ is extraordinary, because: **(1)** It is a ***large, bulk elastic anomaly affected by only a few per cent Cr substitution***. **(2)** NVTE is generally smoothly varying with temperature, found only in a handful of materials where it is attributed to highly anharmonic vibrational modes **[18-21]**. In contrast, the ***"two-step"***

***NVTE we observe for $Ca_2Ru_{1-x}Cr_xO_4$ arises from consecutive, discrete transitions to orbital and magnetic order***, a phenomenon never found before. **(3)** Nearly zero volume thermal expansion is observed over the wide (100 K) interval spanning $T_C$ and $T_{MI}$, which constitutes a strong "Invar effect".

Single crystals of $Ca_2Ru_{1-x}Cr_xO_4$ with $0 \le x \le 0.135$ were grown using a NEC floating zone optical furnace; technical details are described elsewhere **[22].** The structures of single-crystal $Ca_2Ru_{1-x}Cr_xO_4$ samples were refined at various temperatures between 90 and 295 K using a Nonius-Kappa CCD single-crystal X-ray diffractometer (XRD) with sample temperature set using a nitrogen stream. Structures were refined by full-matrix least-squares using the SHELX-97 programs **[22, 23]**. All structures were strongly affected by absorption and extinction, and the data were corrected for anisotropic absorption by comparison of symmetry-equivalent reflections using the program SADABS **[24]**. Detailed structural results and analysis will be published elsewhere **[22]**. The Cr concentration x was determined by energy dispersive X-ray (EDX) analysis. Measurements of magnetization M(T,H) and electrical resistivity $\rho$(T) for T < 400 K were performed using either a Quantum Design PPMS or MPMS. *All measurements (XRD, EDX, M(T,H) and $\rho$(T)) were performed on the same single crystal for each composition x to ensure consistency* (single-crystal XRD requires a small piece taken from each single crystal studied).

**Figure 1** illustrates some overall trends observed in our structural data for $Ca_2Ru_{1-x}Cr_xO_4$ ($0 \le x \le 0.135$) single crystals. Although Cr doping generally preserves the low-T orthorhombic symmetry (*Pbca*), it reduces and eventually suppresses the orthorhombic distortion (e.g., [**b**–**a**] = 0.247 Å for x = 0, but [**b**–**a**] = 0.018 Å for x =

0.135 at T = 90 K; see **Fig. 1b**) **[22]**. The weakening structural distortion is accompanied by relaxation of the Ru-O1-Ru bond angle $\theta$ and elongation of the $RuO_6$ octahedra (i.e., the Ru-O2 bond distance), as shown in **Fig. 1c**. More important, Cr doping causes the unit cell volume V to expand as T is lowered from 295 K to 90 K for $0.032 \leq \mathbf{x} < 0.13$, as shown in **Fig. 1d**.

The first-order MI transition $T_{MI}$ is definitively marked by a jump in the basal plane resistivity $\rho_{ab}$, as shown in **Fig. 2a**; $T_{MI}$ also drops rapidly with increasing x, which we will show is closely related to the structural relaxation. Remarkably, the first-order transition at $T_{MI}$ is suppressed to zero at a critical composition $x_{cr} = 0.135$, ***simultaneous with both*** the disappearance of the orthorhombic distortion (see **Fig. 1b**), and the return of the thermal expansion $\Delta V/V$ to normal behavior (see **Fig. 1d)**. The onsets of orbital order and NVTE are slightly different chiefly because of a strong hysteresis effect due to the first-order transition **[22]**. ***Clearly, the anomalous NVTE is strongly coupled to the onset of orbital ordering.***

The stability of the AFM state is also critically dependent upon both the Ru-O1-Ru bond angle $\theta$ and a concomitant elongation of the apical Ru-O2 distance, which undergo similar, finite shifts at only modest Cr doping levels (**Fig. 1c**). In particular, the magnetic critical temperature $T_C(x)$ slowly increases and peaks at 130 K near x = 0.067. Moreover, the positive Curie-Weiss temperature $\theta_{CW}$ estimated from high-T fits of magnetic data increases with x, indicating an increasing FM character of the volume-averaged exchange interaction (see **Fig. 2b** inset). Given the low saturation moment (~ 0.04 $\mu_B$/f.u. for x = 0.032) and the strong magnetic field dependence of M(H) shown in **Fig. 2c**, the FM behavior is likely a result of spin canting in an otherwise collinear AFM spin arrangement. The $T_C$ peak is followed by a strong downturn and suppression of magnetic order near the critical composition $x_{cr}$ = 0.135 (shown in **Fig. 2b**), ***simultaneous*** with the suppressions of orbital order and NVTE.

The close correlation of NVTE with orbital and magnetic order is documented by data for a representative composition x = 0.067 (see **Fig. 3**), which illustrates the central findings of this work: **(1)** Strong, negative ***linear*** expansion occurs not only along the **b**-axis, but also along the **a**-axis for x > 0 (**Fig. 3a**), which gives rise to the much rarer case of NVTE **[25]**. **(2)** Discontinuities in the **a**-, **b-** and **c**-axis lattice parameters signal a first-order phase transition from a high-T tetragonal, to a low-T orthorhombic phase at 210 K (**Fig. 3a**). **(3)** V abruptly expands by ~ 0.66% with decreasing T near 210 K, and expands again by ~ 0.2% at $T_c$ = 130 K, but changes only slightly between these two temperatures (**Figs. 3b** and **3c**).

The extraordinary, strong dependence of the structural, magnetic and electrical properties of $Ca_2Ru_{1-x}Cr_xO_4$ on Cr content can be explained as follows. The drastic decrease in $T_{MI}$ observed with increasing Cr doping in $Ca_2Ru_{1-x}Cr_xO_4$ closely tracks the rapidly weakening orthorhombicity, as well as the reduced tilt and elongation of $RuO_6$ octahedra (**Figs. 1b and 1c**), as x increases; and the eventual disappearance of $T_{MI}$ is concomitant with vanishing orthorhombicity for $x_{cr}$ = 0.135 (**Fig. 2a**). Indeed, it is predicted that the orbital ground state is governed by $d_{xy}$ ferro-orbital order that is stabilized by the Jahn-Teller distortion **[12]**; Cr doping readily weakens and eventually removes the Jahn-Teller distortion, thus the existence of $d_{xy}$ orbital order and $T_{MI}$. Moreover, the increasing Ru-O2 bond distance and Ru-O1-Ru bond angle (**Fig. 1c**) destabilize the collinear AFM state **[26]**, which, in turn, lead to weak FM behavior and spin canting. A competition between AFM and FM couplings persists to x < 0.135, but Tc decreases rapidly as x increases above 0.06 (**Fig. 2b**). The fact that NVTE does not occur for x = 0 underscores how Cr doping softens the lattice and "unlocks" strongly buckled $RuO_6$ octahedra **[22]**, allowing both the **a**- and **b-**axis to expand while preserving the structural symmetry. Consequently, V abruptly expands on cooling just below $T_{MI}$, where orbital ordering occurs, and further expands at $T_C$. Higher Cr doping relaxes the orthorhombic distortion that, via a highly unusual spin-lattice coupling, weakens the AFM state, and results in an extraordinary increase in volume on cooling.

The principal conclusion we draw from our observations (**Fig. 3**) is that the unique NVTE anomalies in $Ca_2Ru_{1-x}Cr_xO_4$ are ***directly associated with the onsets of orbital ordering at $T_{MI}$, and magnetic ordering at $T_C$***; and ***between $T_{MI}$ and $T_C$ V remains essentially constant***, ***an Invar effect*** (e.g., the slight or nearly zero thermal expansion first observed in certain Ni-Fe alloys **[27]**). A phase diagram given in **Fig. 4** shows that the first-order ($T_{MI}$) and second-order ($T_C$) transition lines separate three states that meet at a tricritical point. The coexisting orbital and magnetic orders "melt" when both $T_{MI}$ and $T_C$ disappear near $x = 0.12 \pm 0.01 \approx x_{cr}$, implying that x is a critical Cr concentration.

NVTE is seen in only a handful of materials such as $ZrW_2O_8$ and $ZnCr_2Se_4$, where it is primarily associated with transverse thermal motion of oxygen or highly anharmonic vibrational modes **[17-20]**; the latter effect is related to spin canting and a spin-lattice coupling at low temperatures **[21]**; but in these cases, no anomalies in other physical properties (e.g., a MI transition) are observed. In contrast, the unusual "two-step" temperature dependence of V and the simultaneous suppressions of magnetic and orbital order and orthorhombic distortion with dilute Cr doping indicate the existence of a novel coupling between spin, orbital and lattice degrees of freedom in $Ca_2Ru_{1-x}Cr_xO_4$.

The conspicuously wide separation between $T_{MI}$, on the one hand, and $T_N$ or $T_C$ on the other, for the pure (x = 0) and more dilute Cr-doped samples poses the surprising possibility that spin-orbit coupling has a more subtle influence on the physical properties of $Ca_2Ru_{1-x}Cr_xO_4$ than is commonly anticipated for other ruthenates **[7, 8]**. The nearly constant V observed between $T_{MI}$ and $T_C$ is a new manifestation of the technologically important Invar effect, and reinforces the conclusion that the strong NVTE in $Ca_2Ru_{1-x}Cr_xO_4$ is closely coupled to the onsets of orbital and magnetic order.

The ***negative volume thermal expansion via orbital and magnetic orders*** found in $Ca_2Ru_{1-x}Cr_xO_4$ suggests an exotic ground state, as well as a paradigm for functional materials with highly unusual thermal expansion and electronic characteristics.

This work was supported by NSF through grants DMR-0552267, DMR-0856234 (GC) and EPS-0814194 (GC, LED), and by DoE through grants DE-FG02-97ER45653 (LED) and DE-FG02-98ER45707 (PS).

*To whom correspondence should be addressed; email: cao@uky.edu

**Figure Captions:**

**Fig. 1. (a)** Temperature dependence of the **a**- and **b**-axis lattice parameters, and the ab-plane resistivity $\rho_{ab}$ (right scale) for x = 0. **(b)** Cr concentration (x) dependence of the **a**-, **b**- and **c**-axis lattice parameters (right scale) for T = 90 K. **(c)** Ru-O1-Ru bond angle and the Ru-O2 bond distance (right scale) for T = 90 K. **(d)** Unit cell volume V for T = 90 K and 295 K, and thermal expansion ratio [V(295K)-V(90K)]/V(295K) (right scale) for $0 \leq x \leq 0.135$. **Fig. 1a Inset**: Schematics of orbital order and disorder. **Fig. 1c Inset:** Schematics of the distorted Ru-O1-Ru bond angle $\theta$ and a $RuO_6$ octahedron.

**Fig. 2.** Temperature dependences of **(a)** the ab-plane resistivity $\rho_{ab}$, **(b)** the magnetic susceptibility $\chi_{ab}$ at applied field $\mu_o H$ = 0.5 T for $0 \leq x \leq 0.135$, and **(c)** the isothermal magnetization M(H) at 100 K for representative compositions x = 0 and 0.032. **Fig. 2b Inset:** $T_N$, $T_C$ and Curie-Weiss temperature $\theta_{CW}$ vs. x.

**Fig. 3.** For x = 0.067, temperature dependences of: **(a)** lattice parameters **a**-, **b**- and **c**-axis (right scale), **(b)** unit cell volume V and thermal expansion ratio $\Delta V/V$ (right scale), and **(c)** ab-plane resistivity $\rho_{ab}$ and magnetic susceptibility $\chi_{ab}$ at $\mu_o H$ = 0.5 T. The shaded area indicates a region of mixed tetragonal and orthorhombic phases.

**Fig. 4.** The T-x phase diagram summarizing observed phase transitions and phase types. Right scale shows the Cr concentration (x) dependence of the thermal expansion ratio $\Delta V/V$. Note that the hatched region represents negative volume thermal expansion.

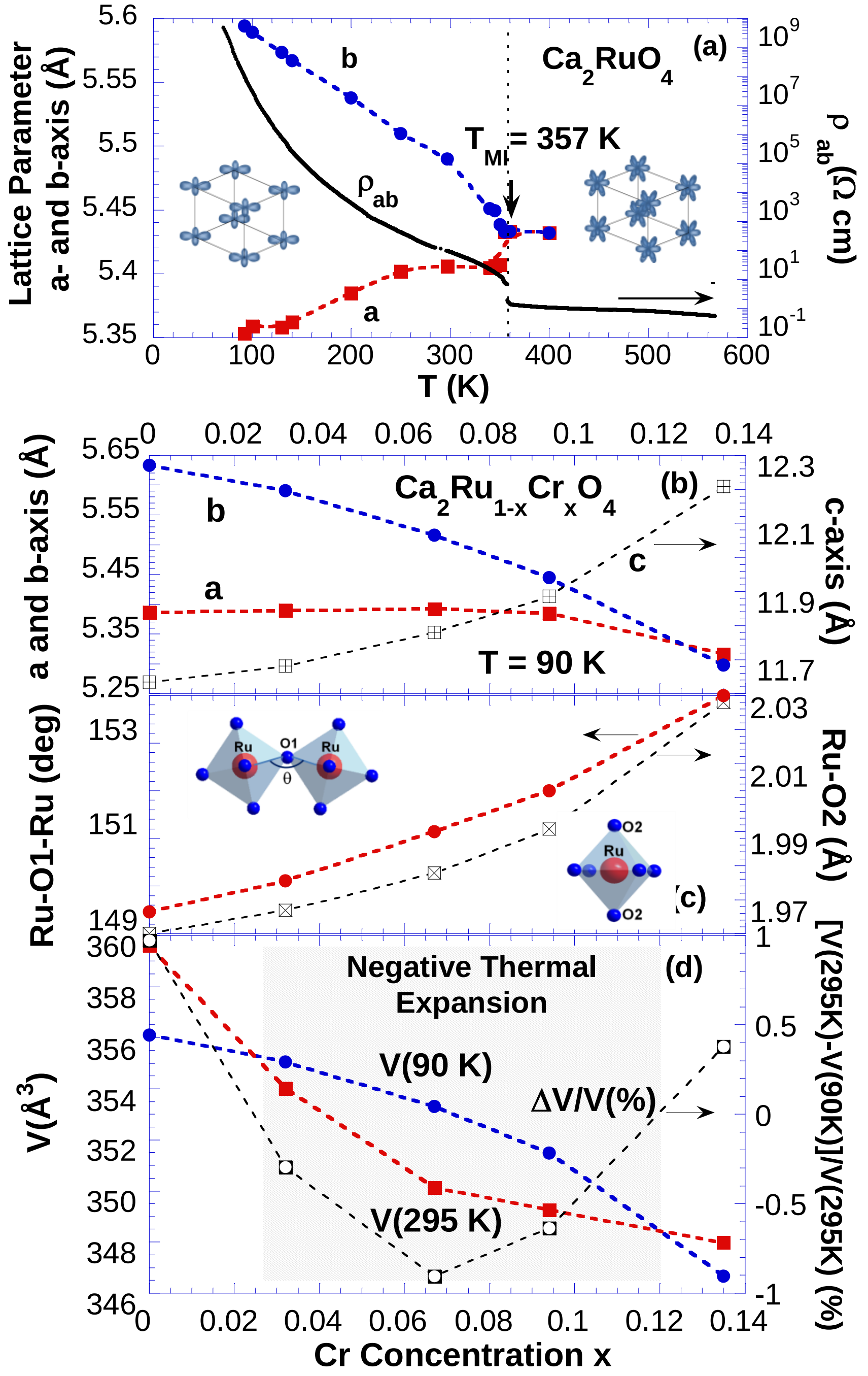

Fig.1

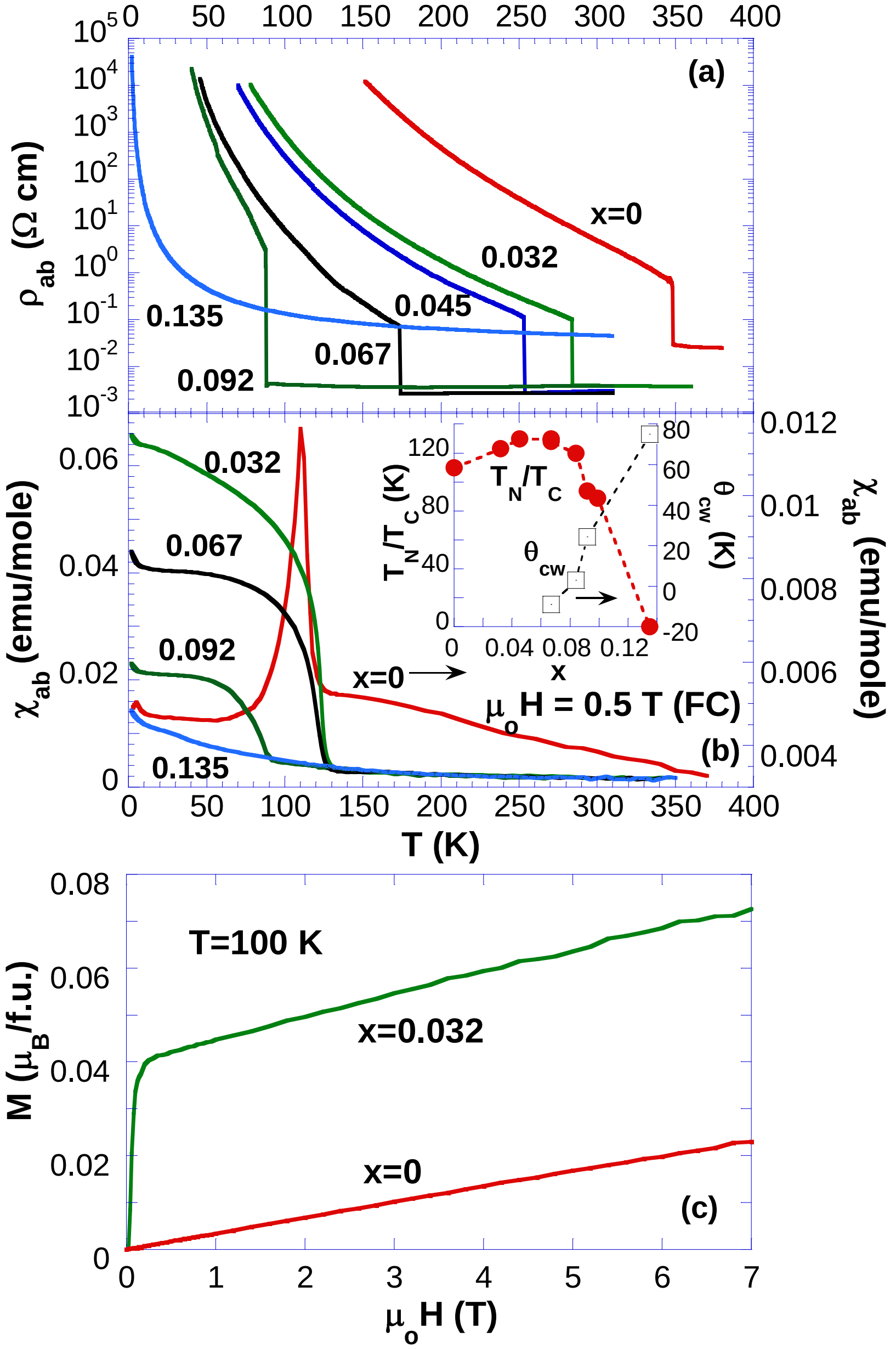

Fig.2

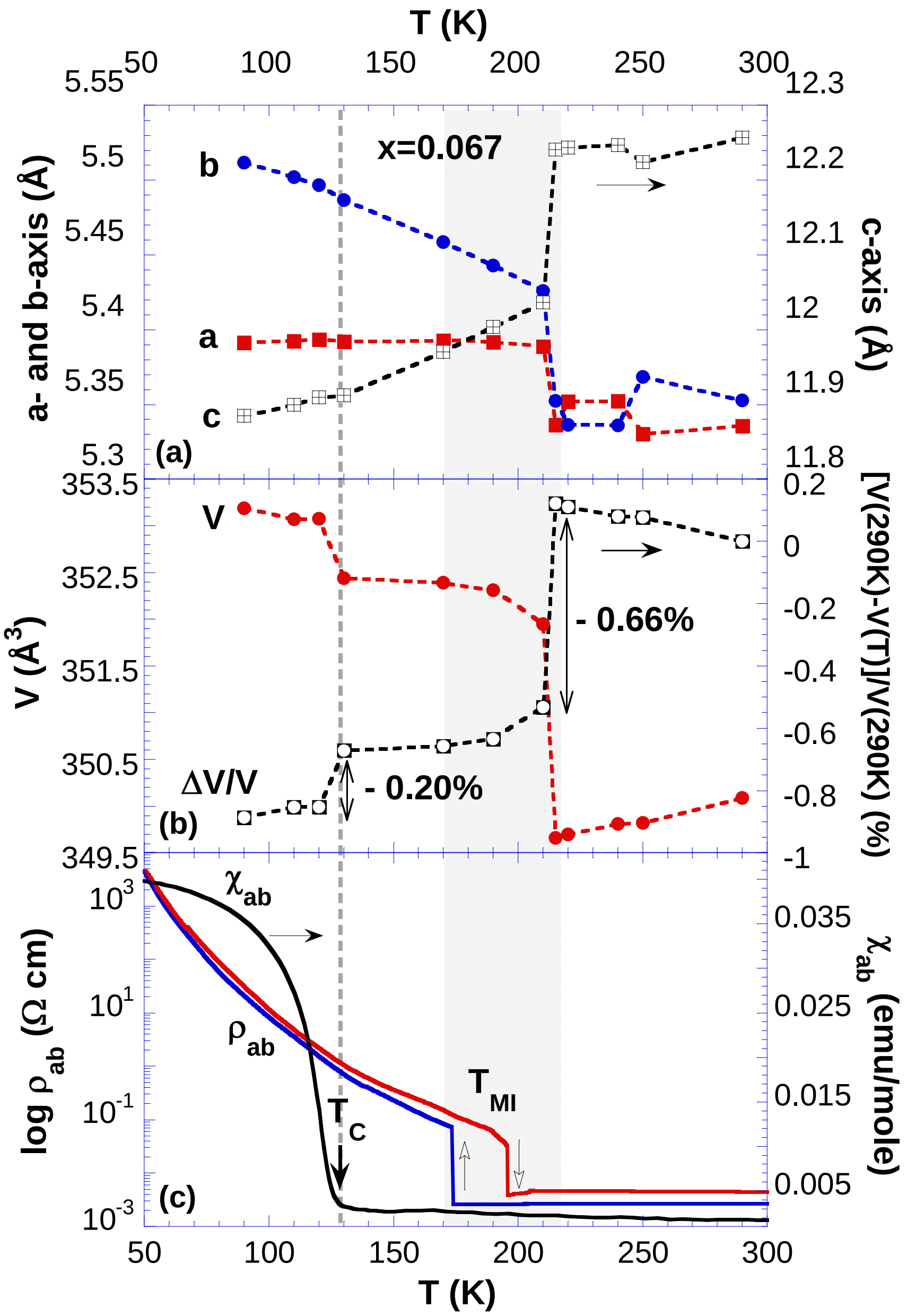

Fig. 3

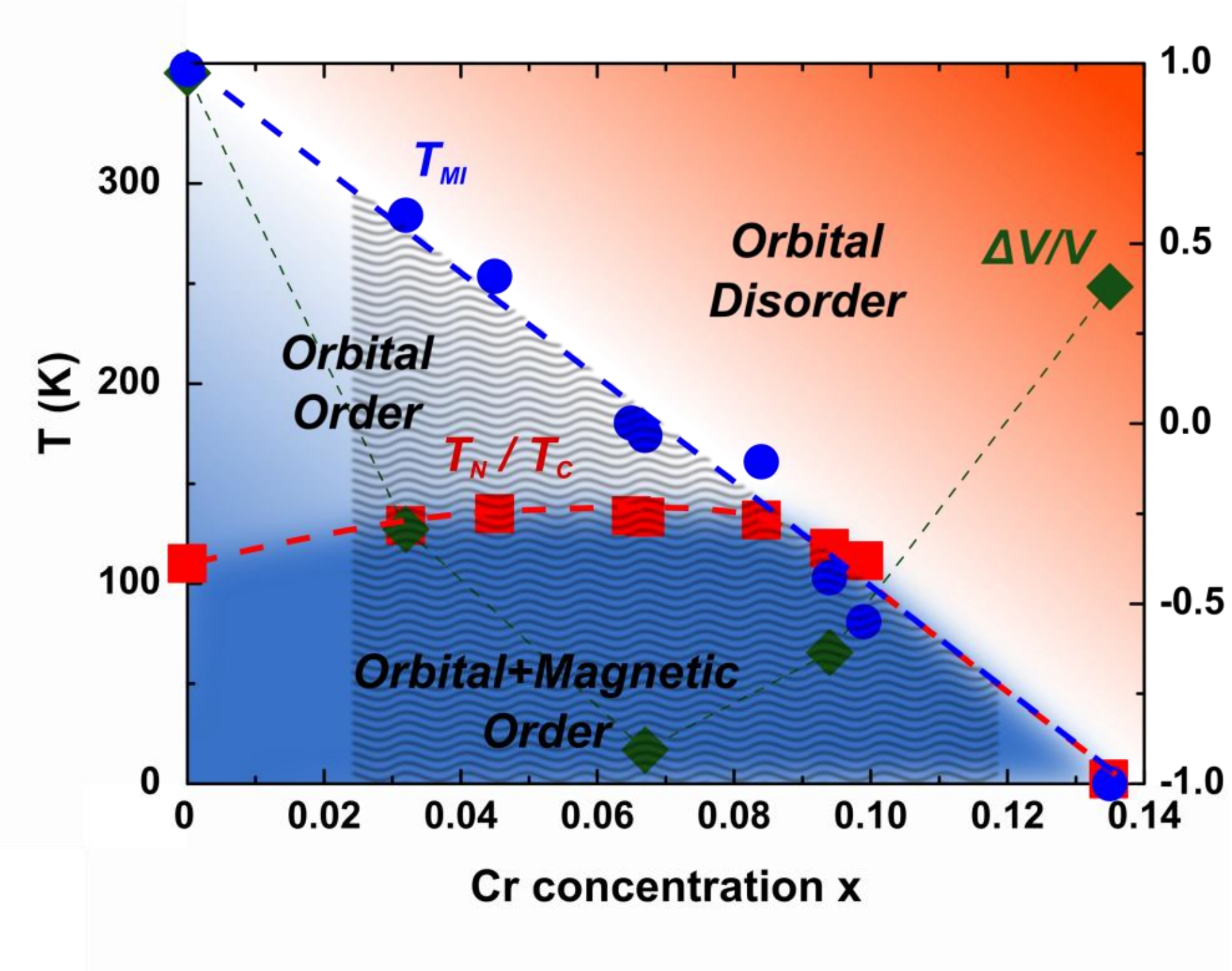

Fig.4